\documentclass[pdflatex,sn-mathphys-num]{sn-jnl}

\usepackage{graphicx}%
\usepackage{multirow}%
\usepackage{amsmath,amssymb,amsfonts}%
\usepackage{amsthm}%
\usepackage{mathrsfs}%
\usepackage{xcolor}%
\usepackage{textcomp}%
\usepackage{manyfoot}%
\usepackage{booktabs}%
\usepackage{algorithm}%
\usepackage{algorithmicx}%
\usepackage{algpseudocode}%
\usepackage{listings}%
\usepackage{xfrac}
\usepackage{graphicx}
 \usepackage{subfig}
\newcommand{\bk}{\mathbf{k}}
\newcommand{\bu}{\mathbf{u}}
\newcommand{\bn}{\mathbf{n}}

\theoremstyle{thmstyleone}%
\theoremstyle{thmstyletwo}%

\theoremstyle{thmstylethree}%

\begin{document}

\title[Fine Structures Roll up in the Flow of Film Boiling at High Density Ratios ]{Fine Structures Roll up in the Flow of Film Boiling at High Density Ratios }


\author*[1]{\fnm{Saeed} \sur{Mortazavi}}\email{saeedm@iut.ac.ir }

\author[2]{\fnm{Iman} \sur{Yaali}}\email{i.yaali@me.iut.ac.ir}


\affil[1,2]{\orgdiv{Department of Mechanical Engineering}, \orgname{Isfahan University of Technology}, \postcode{84156-83111}, \state{Isfahan}, \country{Iran}}




\abstract{Film boiling has practical applications in the current technology including steam power plants, cooling of electronic devices and emergency cooling systems. A finite difference/front tracking method is used to simulate film boiling at high density ratios on a horizontal plate subject to a constant wall heat flux. The grid resolution is relatively high (768 grids per width of the domain). The flow is dominated by Rayleigh-Taylor instability as well as Kelvin-Helmholtz instability. The flow structure includes the roll up of the interface between the gas and liquid. This happens at high density ratio (1000) where the difference between the gas and the liquid velocities  across the interface is large. The jump in tangential velocity is an order of magnitude smaller at a lower density ratio (100). Hence, there is no roll up at lower density ratio. The flow is also influenced by vortex development as a result of the baroclinic term in the vorticity transport equation. The density gradient is large at the interface at high density ratio which tends to increase the baroclinic term. The plot of pressure gradient and density gradient shows that they are not parallel in the roll up regions. As a result, vortices in small scales develop that shed in the gas phase. The plot of the enstrophy with time shows that it shows smooth variation that match for two grid resolutions, however at a specific time enstrophies become spiky, and they depart from each other at two grid resolutions. The spiky behavior of enstrophy is due to vortex shedding in the roll up region.}




\maketitle

\section{Introduction}\label{sec1}

Film boiling has many practical applications in present technology and industries. Steam power plants are the basic application of boiling where water is converted to steam for power generation.  The storage of energy in the form of latent heat has also practical application in power plants. The cooling of electronic devices entails the phase change that applies to boiling process. There are numerous experimental, analytical and numerical efforts in the literature to understand the boiling phenomenon. 
     Boiling process is hampered by very small time and space scales that is very difficult to resolve experimentally. Hence, analytical tools and numerical efforts have been applied to boiling in order to overcome the difficulties encountered. Experimental investigations have resulted empirical correlations that estimate the Nusselt number with a large degree of uncertainty. Analytical efforts have been applied to very simple geometries with simplified assumptions. These include the work by Rayleigh~\cite{rayleigh1917pressure} who obtained an expression for the growth of a vapor bubble. 
     Recently, direct numerical simulation has been applied to study boiling flows. There are a majority of difficulties encountered in simulating boiling flows. The jump in physical properties across the interface, mass transfer, heat absorption and surface tension calculation are common examples that are present in the numerical method. Early attempt in computation of boiling turns back to Welch \cite{welch1995local} who used a moving mesh to solve the momentum and energy equations in two dimensions. Juric and Tryggvason \cite{juric1998computations} used a finite difference front tracking method to simulate film boiling on a flat surface subject to a constant heat flux. Esmaeeli and Tryggvason \cite{esmaeeli2004computations} developed a new algorithm to simulate film boiling using Front-Tracking method.
     Khorram and Mortazavi \cite{khorram2022direct} investigated film boiling on a horizontal periodic surface using front tracking through direct numerical simulation (DNS). They introduced a new topology changing algorithm for breakup and coalescence of bubbles. In their method, the bubble release cycles were repeated several times, which lead to prolongation of the simulations. Mortazavi \cite{mortazavi2022toward} performed simulations of film boiling at high density ratios (2000). The microstructure of the flow was studied that included the circulation zones and a vortex pair developed under the bubble root. At high density ratio, the flow develops a wide neck at the bubble root where a vortex pair is generated. 
      Boiling flows deal with Rayleigh-Taylor instability since gravity is present, and liquid stands on the top of the vapor. Also, since the liquid velocity is different from the gas velocity across the interface, they are subject to Kelvin-Helmholtz instability. It turns out that the interface is unstable to small perturbations in Kelvin-Helmholtz instability as the linear stability theory implies. The interface rolls up as a result of the instability as has been observed by several investigators (see Tryggvason et al \cite{tryggvason1991fine}). Several investigators tried to examine the evolution of a vortex sheet, and ended up with the formation of a singularity developed during the evolution. It was further shown that the presence of surface tension suppresses the formation of singularities, and short wavelengths instability is removed (see Siegel \cite{siegel1995study}). Other investigators used the Navier-Stokes equations and studied the evolution of a vortex sheet. They showed that the perturbed shear layer turns to a rows of vortices (see Patnaik et el, \cite{patnaik1976numerical} and Crocos and Sherman \cite{corcos1984mixing}). Tauber et al \cite{tauber2002nonlinear} examined the behavior of the interface between two immiscible fluids when there is a jump in the tangential velocity across the interface. They observed that fingers of interpenetrating fluids are formed along the interface. 
     The present study is the extension of the work by Mortazavi \cite{mortazavi2022toward} on film boiling at high density ratios. High density ratios were examined in that work, however the simulations were performed at relatively moderate grid resolutions. Here we select high grid resolutions so that fine structures where small vortices are formed are captured. The interface rolls up on the sides of the front due to the Kelvin-Helmholtz instability, and develops vortices as a result of the baroclinic term in the vorticity transport equation. 
      Here, the interface in film boiling at large density ratios (1000) entails a shear layer due to the difference between the liquid and gas velocities across the interface. As a result, the interface rolls up due to shear layer instability. This does not happen at a lower density ratio (100) since the difference in velocity across the interface is an order of magnitude smaller. Also, the baroclinic term in the vorticity transport equation becomes large on the interface at high density ratios. Mathematically, this term reads as: $\sfrac{\nabla\rho\times\nabla p}{\rho^2}$.

      It is evident that this term is directly proportional to the jump in density across the interface. Taking the liquid density as a reference density (for example), this term is proportional to the density ratio existing in the flow. This generates vortices in the roll up region where this term shed vortices in the gas phase. The phenomenon is observed through numerical simulations at relatively high grid resolution. A stability analysis is also performed on the flow that shows the growth rate is positive, and the flow is unstable to small perturbations. The stability analysis requires the jump in tangential velocity across the interface. This is estimated by calculating the average gas and liquid velocities from numerical simulations. The roll up of the interface entails some gas patches the may separate from the rest of the gas and suspend inside the liquid. This may change the flow structure in the neighborhood of the interface. As a result, the flow mixing and liquid evaporation will be affected by these small gas patches around the interface. The novelty of the present work included the new morphology of the interface that is changed by shedding of the vertices around the interface. The roll up of the interface affects the flow mixing in these regions that affects the liquid evaporation in these areas.

\section{Problem definition and governing equations}\label{sec2}

We use conservation of mass, momentum and energy to simulate film boiling on a flat horizontal plate. The equations are written in conservative form, and one set of equations are used for both phases with variable physical properties. The surface tension force is added to the momentum equation by a source term that acts at the interface. This is done by a delta function that is non-zero on the interface. The heat absorbed as a result of liquid evaporation is added as a source term to the energy equation. This is also done by a delta function that is non-zero only on the interface. Both the liquid and the gas are assumed to be incompressible. Thus, the flow is divergence free in these phases. However, due to the phase change and liquid evaporation, there is a volume change in the liquid that evaporates.  In fact, the conservation of mass reduces to: 
\begin{align}\label{Eq1}
          \nabla\ldotp \bu =0	
          \end{align}

except at the interface where phase change occurs. 
      The Navier-Stokes equations are solved in conservative form with variable physical properties: 

\begin{align}
          \frac{\partial \rho \textbf{u}}{\partial t}+\nabla \ldotp(\rho \textbf{uu})=-\nabla P+\rho {\textbf{g}}+\nabla\ldotp \mu (\nabla \textbf{u}+\nabla \textbf{u}^T )
          +\int_A \gamma \bk \bn \delta (\textbf{x}-\textbf{x}_f )dA_f 
        \end{align}
where $\gamma$ is the surface tension coefficient, $\bk$ is the curvature for two-dimensional flow, and twice the mean curvature for three-dimensional flow. The subscript $f$ stands for the interface position. $n$ is an outward unit normal vector towards the gas, $A$ is the area of the interface and $x$ represents the position vector where the equation is evaluated. The energy equation is considered in conservative form with variable physical properties that is valid for both phases:
    
\begin{align}\label{Eq3}
   \frac{\partial \rho C T}{\partial t}+\nabla \ldotp\left(\rho C\bu T\right)=\nabla\ldotp\left(k\nabla T\right)-\left(1-\left(C_g-C_{\ell}\right)\frac{T_{\mathsf{Sat}}}{h_{fg}}\right)\int_{A}\delta\left(x-x_f\right){\dot{q}}_fdA_f
  \end{align}         

  where $C$ is the constant pressure specific heat, $k$ is the thermal conductivity, T is the temperature, $h_fg$ is the latent heat of vaporization and $T{_sat}$ is the saturation temperature. The heat absorbed as a result of liquid evaporation is added as a source term that acts at the interface. The source term in the energy equation is balanced by the difference in the heat conduction on both sides of the interface: 
  \begin{align}\label{Eq4}
    {\dot{q}}_f=k_g\left.\frac{\partial T}{\partial n}\right|_g-k_l \left.\frac{\partial T}{\partial n}\right|_l\ 
    \end{align}

    This is basically the jump condition that is satisfied on the interface. It is assumed that the interface is at the saturation temperature corresponding to the system pressure. The evaporation rate, i.e., the mass flow rate due to evaporation per unit area is related to the heat absorbed at the phase boundary through: 
    \begin{align}
     {\dot{q}}_f=\dot{m}h_{fg}
     \end{align}

     where $\dot{m}$ is the evaporation rate. If there is phase change inside the flow the divergence of the velocity reduces to:  
     
     \begin{align}\label{Eq6}
     \nabla\ldotp\bu=\frac{1}{h_{fg}}(\frac{1}{\rho_g}-\frac{1}{\rho_{\ell}})\int_{A}\delta\left(x-x_f\right){\dot{q}}_fdA_f
     \end{align}

     The above result is valid throughout the flow, and it includes equation~\eqref{Eq1} as well, since the delta function is zero when it is evaluated away from the interface. Therefore Equation~\eqref{Eq6} replaces Equation~\eqref{Eq1} if there is phase change inside the flow. The reader is referred to the article by Esmaeeli and Tryggvason \cite{esmaeeli2004computations} for a detailed derivation of equation~\eqref{Eq6}.    
     The plate is subject to a constant heat flux and a film of vapor covers the plate initially. An outflow boundary condition is assumed at the upper boundary. The lateral boundaries are assumed to be periodic in order to avoid the effect of any solid boundaries in the horizontal direction. The height of the computational domain is 1.5 times the domain length.  A symmetric initial perturbation is assumed for the front:
     
    \begin{align}
     y = y_c+\varepsilon \left(\cos{\frac{2\pi x}{X_{\ell}}}\right)
      \end{align}
      where $y_{c}=0.2X_{\ell}$ is the initial interface height, 
      $\varepsilon =-0.07X_{\ell}$, is the perturbation amplitude and $X_{\ell}$ is the domain length. 
\section{Numerical Method }\label{sec3}



The governing equations for multi-phase medium are solved by a finite difference/front tracking method developed originally by Unverdi and Tryggvason \cite{unverdi1992front}. The method did not include phase change across interface. Juric and Tryggvason \cite{juric1998computations} made further improvement of the method by adding the energy equation and phase change. Also, Esmaeeli and Tryggvason \cite{esmaeeli2004computations} developed a new algorithm to include phase change, and applied the method to film boiling. Here we use the same algorithm as Esmaeeli and Tryggvason \cite{esmaeeli2004computations} to simulate film boiling at high density ratios. The code has been developed by the author independently, and applied to several phase change problems. (see Mortazavi \cite{mortazavi2022toward}). The algorithm has been elaborated in detail by Esmaeeli and Tryggvason \cite{esmaeeli2004computations}. We briefly discuss the method by pointing out the important issues. The familiar projection method is used the solve the Navier-Stokes equations. A Poisson equation is solved at every time step to find the pressure. The divergence of the velocity is zero everywhere except at the interface. Equation~\eqref{Eq6} is used to obtain the divergence of the velocity around the location of the interface. According to equation~\eqref{Eq6} divergence of the velocity is zero except in regions around the interface where phase change occurs. Therefore, an extra term is added to the Poisson equation for pressure that includes the divergence of the velocity field.
The source term in the energy equation is evaluated using equation~\eqref{Eq4}. To calculate this term a probe is used at the interface that is normal to the interface and points towards the gas or liquid. Then the temperature gradient normal to the interface is found both in the liquid phase and the gas phase. The source term is then distributed on the Eulerian grid where the governing equations are solved. A Peskin distribution \cite{peskin1977numerical} is used to distribute the source term on the Eulerian grid. The same procedure is used to calculate the divergence of the velocity field. 
The convective terms are discretized by QUICK algorithm which is third order accurate in space. The diffusion terms are discretized using the normal central differencing. The integration in time is done by a predictor corrector scheme that is second order accurate. The Poisson equation for pressure is solved by a multi-grid method developed by Adams \cite{adams1989mudpack}. 
Pool boiling is considered in the present problem. Thus, we are encountered with a stagnant ambient liquid. As a result, forced convection is not present. This affects the interface motion when dealing with boiling. The interface is moved using Lagrangian description for the interface:

  \begin{align}
 \frac{d{\bar{\textbf{x}}}_f}{dt}=\left(\frac{1}{2}\left({\bar{U}}_{\ell}+{\bar{U}}_{g}\right)\ldotp \bar{\textbf{n}}-\frac{\dot{q}_{f}}{2h_{fg}}\left(\frac{1}{\rho_{\ell}}+\frac{1}{\rho_{g}}\right)\right) \bar{\textbf{n}}_{f}
  \end{align}

   The interface is only moved by the normal component of the velocity. The first term is basically the convective effect. The second term is a result of phase change or evaporation. It has only one component which is along the local normal to the interface. The first term is calculated by interpolating the velocity from the Eulerian grid using Peskin interpolation \cite{peskin1977numerical}. The second term is known once the source term in the energy equation is evaluated
\section{Dimensionless parameters}\label{sec4}
    Dimensionless parameters in the flow are derived from the physical properties of the gas and liquid phase. The Morton number is defined as: $Mo=\frac{\mu_{\ell}^{4}g}{\gamma^{3}\rho_{\ell}}$, the Capillary number is: $Ca=\sfrac{C_{\ell}T_{\mathsf{Sat}}\gamma}{\rho_{v}h_{fg}\ell}$, 
    the Prandtl\ number is: 
    $Pr=\sfrac{\mu c_{\ell}}{k_l}$\ 
, the dimensionless
heat flux at the lower heated wall is: 
$q_w^\star\ = \frac{q_w}{\frac{k_l}{c_l} \frac{h_{fg}}{l}\ }$ .
 The subscript $\l$ and $v$ refer to the liquid and gas phase respectively, $\gamma$ is the interfacial tension, ${h_{fg}}$,\
     is the latent heat and $\ell$
     is a characteristic length defined as: 
     $\ell =\left(\sfrac{\mu_{\ell}^{2}}{g\rho_{\ell}^{2}}\right)^{\sfrac{1}{3}}$. 

\noindent It is assumed that the interface is at the saturation temperature, $T_{\mathsf{Sat}}$. The most unstable wavelength of inviscid Rayleigh-Taylor instability is defined as 
       ${\lambda_{d_{2}}=2\pi\left(\sfrac{3\gamma}{g(\rho_l-\rho_g)}\right)}^{\sfrac{1}{2}}$ and the Nusselt number is defined as 
      $Nu\ =\frac{q_w \ell}{k_g (T_w-T_{\mathsf{Sat}})}$
      which is a dimensionless wall temperature. Also, the physical property ratios are common dimensionless parameters: 
      $r={\rho}_{\ell}/\rho_{g}$,  $\mu_{\ell}/\mu_{g}$,  $c_{\ell}/c_{g} $, $k_{\ell}/k_{g}$. A characteristic time can be defined as:
      $t_c=\left(\sfrac{\ell}{g}\right)^{\sfrac{1}{2}}$, and a characteristic velocity is defined as:
      $u_c=\left(\lambda_{d_{2}}g\right)^{\sfrac{1}{2}}$. The domain length is assumed as two times $\lambda_{d_{2}}$ as is taken in the past work by Mortazavi~\cite{mortazavi2022toward}. Table~\ref{tab1} indicates dimensionless parameters used in the current work that correspond to hydrogen as is taken by Juric and Tryggvason \cite{juric1998computations}. It should be pointed out that here the interfacial tension affects the flow through the capillary number, and also the domain length as taken based on $\lambda_{d_{2}}$. Any other choice of the interfacial tension influences the simulation through these parameters. The interfacial tension is included in the simulations through the momentum equation.
        \begin{table}[h]
          \caption{Dimensionless parameters in the present study}\label{tab1}
       \begin{tabular*}{\textwidth}{@{\extracolsep\fill}lccccccc}
       \toprule%
         $\sfrac{\rho_l}{\rho_g}$ & $\sfrac{\mu_l}{\mu_g}$ & $\sfrac{k_l}{k_g}$ & $ q_w^\star$ & $Pr$ & $\sfrac{c_l}/{c_g}$ & $Mo$ & $Ca$  \\
         \midrule
        100,1000  & 40.0 & 10.0 & 0.025 &  1.0 & 1.0 & 0.1 & 0.0\\
         
         \botrule
          \end{tabular*}
        
        \end{table}
        The capillary number is set to zero by taking the saturation temperature equal to zero as is taken by Juric and Tryggvason \cite{juric1998computations} in some cases. However, the interfacial tension is not zero in the simulations. The saturation temperature affects the source term in the energy equation (equation~\eqref{Eq3}), that depends on it. Otherwise, the numerical value of the saturation temperature does not affect the simulations as long as the difference in the temperature plays the important role. For the case considered here since the specific heats of the gas and the liquid are the same the saturation temperature does not play any role (the term that includes the saturation temperature in equation~\eqref{Eq3} is identically zero). For validation of the numerical method at high density ratios see Mortazavi \cite{mortazavi2022toward}. The computational code has been validated, and benchmark tests were performed to validate the numerical method at high density ratios. The reader is referred to the work by Mortazavi \cite{mortazavi2022toward} for benchmark tests. 
\section{Results and Discussion }\label{sec5}

\begin{figure}[H]
	     	\centering
	     	{\includegraphics[width=6cm]{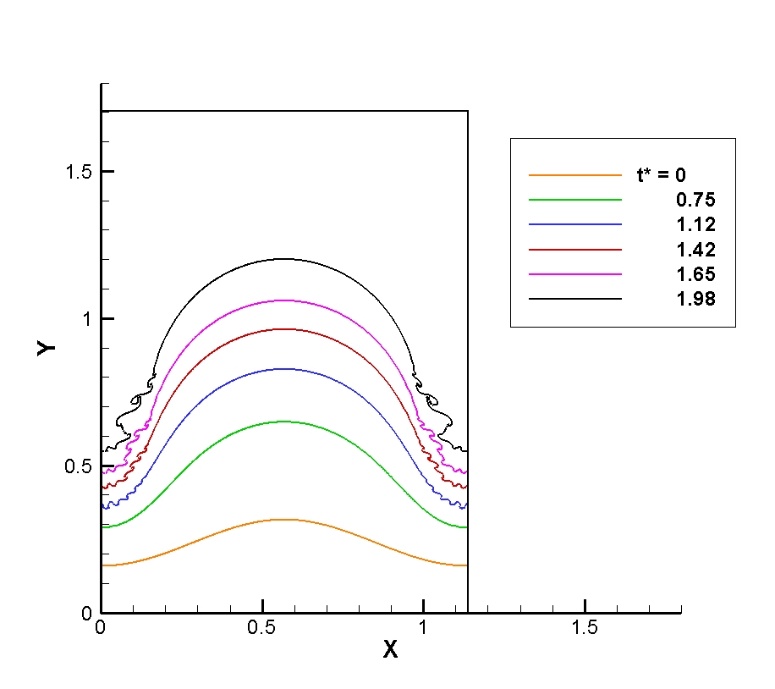}}
	     	\qquad
	     	\caption{Typical front positions for film boiling at a high density ratio (1000). The grid resolution is $512\times768$.}
	     	\label{Fig:high density}
	     \end{figure}

     Simulations are presented at high grid resolution to capture fine structures due to small vortices developed as a result of large density gradient in the flow. Vortices are formed due to the baroclinic term in the vorticity transport equation. This term depends on the density gradient as well as the pressure gradient which presumes large values when the density ratio is large in the simulations (1000). Specifically, these two gradients obtain large values in the neighborhood of the interface. The domain length is two times the most unstable wavelength of Rayleigh-Taylor instability, $\lambda_{2d}$. This length has been taken in the previous work by Jouric and Tryggvason~\cite{juric1998computations}, and Mortazavi~ \cite{mortazavi2022toward}. As has been shown by Mortazavi~\cite{mortazavi2022toward}, increasing the domain length larger than $\lambda_{2d}$ amplifies the instability in the flow, and the interface grows faster with less resistance against gravity. The lateral distance is one and a half times the length. Typical front shapes are plotted in Fig.~\ref{Fig:high density} at different times for a simulation with $512\times768$ grid resolution. A parametric study of the problem has been performed at high density ratios in the work by Mortazavi \cite{mortazavi2022toward}. The effect of the Moton number, dimensionless wall heat flux and the density ratio has been performed in detail in reference \cite{mortazavi2022toward}. 

\begin{figure}[H]
	     	\centering
	\subfloat{\includegraphics[width=5cm]{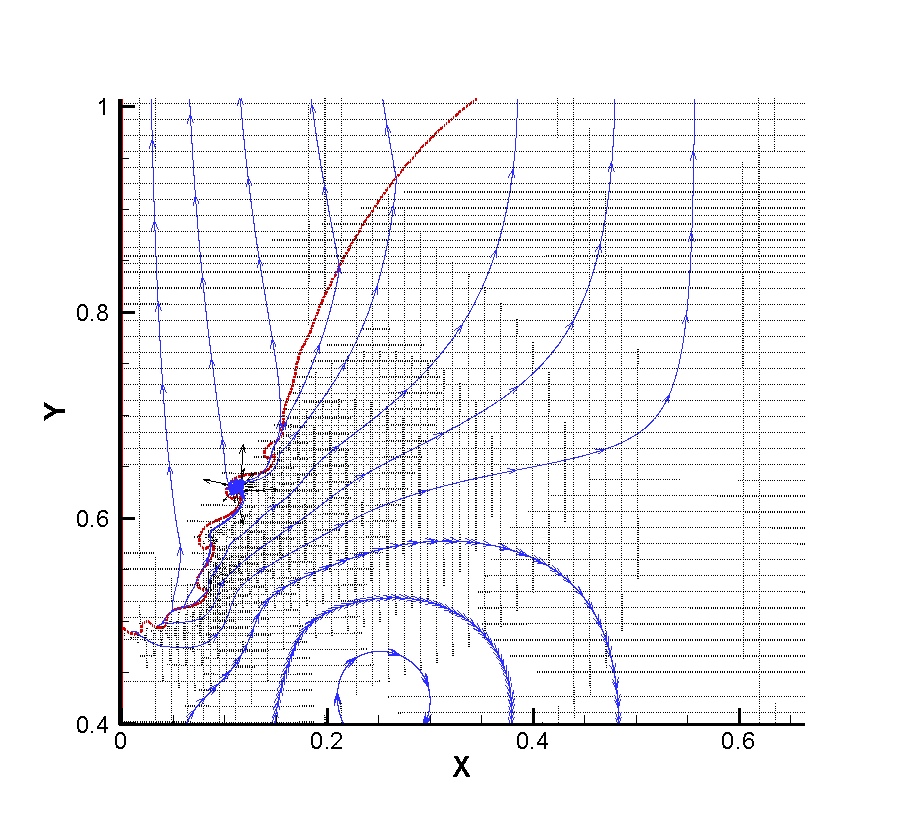}}
	     	\qquad
\subfloat{\includegraphics[width=5cm]{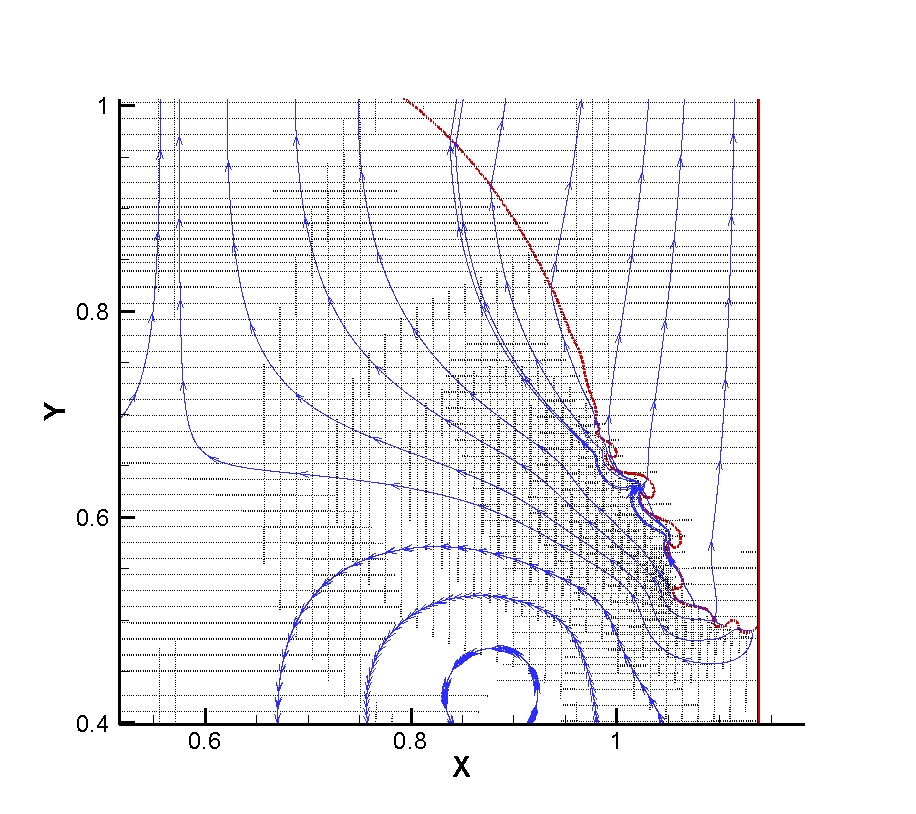}}
                 \qquad
    \subfloat{\includegraphics[width=5cm]{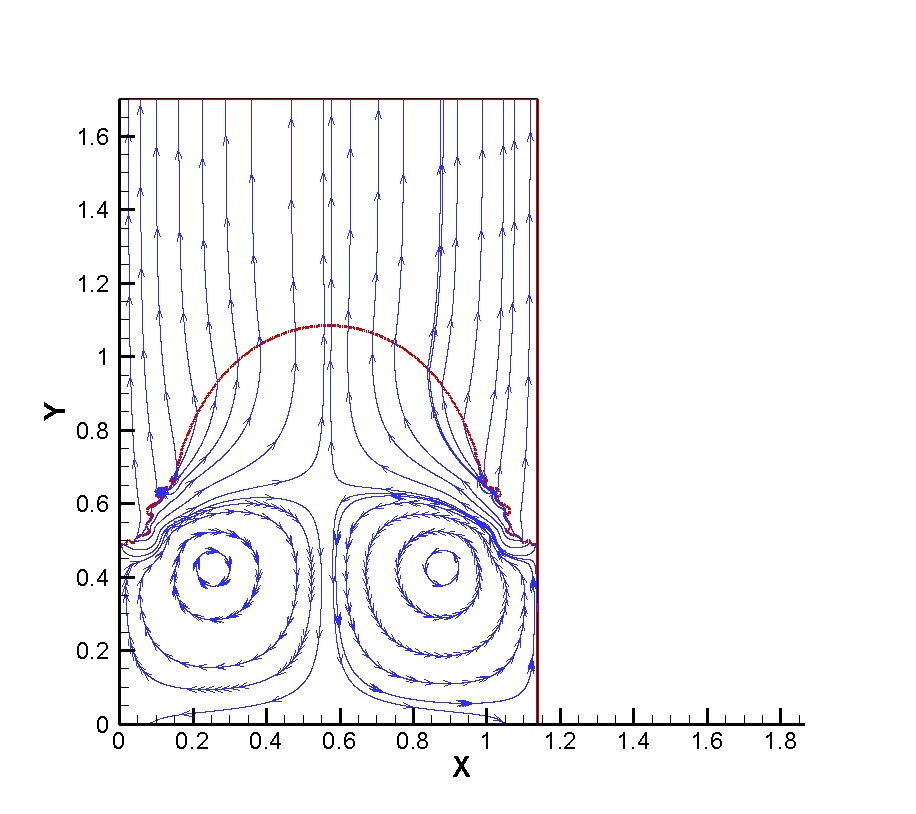}}
                 \qquad
    \subfloat{\includegraphics[width=5cm]{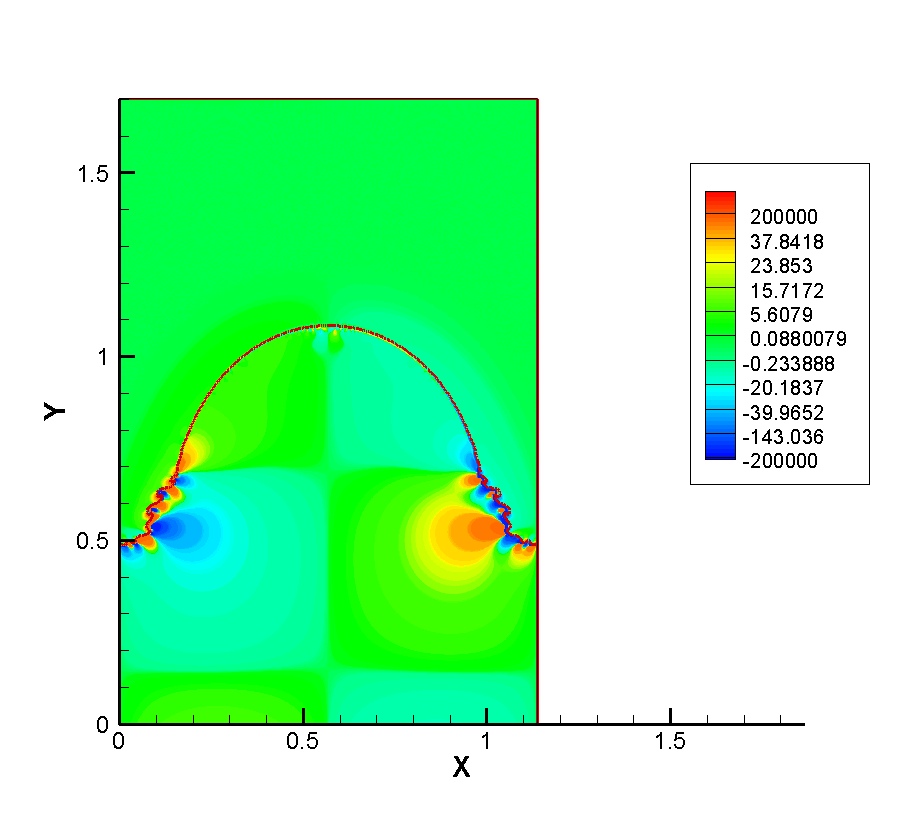}}
	     	\caption{The flow structure at time 1.71 for film boiling at a density ratio 1000. Typical vortices appear in the roll up regions in the left and right side of the front where the interface rolls up in these areas. The bottom right frame is a contour plot of vorticity. The grid resolution is $512\times768$.}
	     	\label{Fig:The flow structure}
	     \end{figure}

\begin{figure}[H]
	     	\centering
	     	\subfloat{\includegraphics[width=8cm]{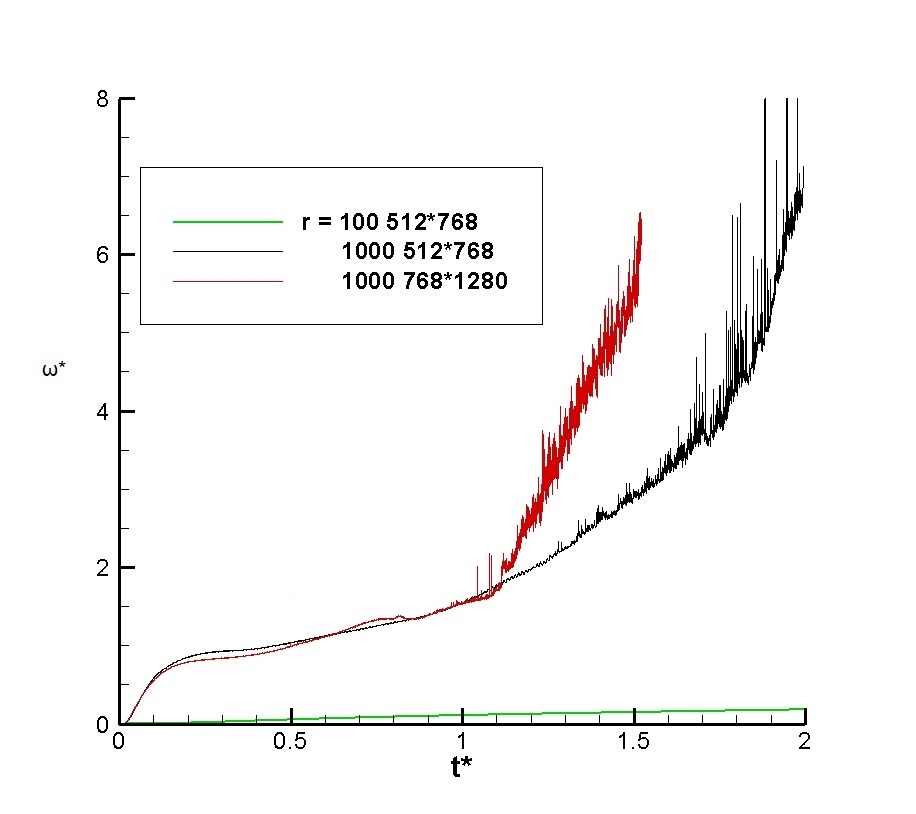}}
	     	\qquad
	     	\caption{Absolute value of vorticity (enstrophy), averaged on the computational domain with respect to time for two density ratios (100,1000). Same trend is observed up to time 1.0 for two grid resolutions at a high density ratio (1000).}
        \label{Fig:9}
    
             \end{figure}

         Fig.~\ref{Fig:The flow structure} indicates the flow structure at dimensionless time 1.71 when the interface has rolled up a few times in the left and right areas. Vortices appear in these areas as indicated by the streamlines in the core of the vortices. The roll up of the interface is due to the Kelvin-Helmholtz instability developed since there is a relatively large difference between the gas velocity and the liquid velocity across the interface. The vortices develop in the roll up region where the baroclinic term gets large at the interface and induces vortices in the flow. These vortices appear in small scales, so the grid resolution should be large enough to capture these vortices. Vorticity is relatively small at initial stages of the simulation as indicated by the absolute value of vorticity in Fig.~\ref{Fig:9}. Vorticity has a smooth variation with time and increases with time as is seen in the figure. This trend changes at time 1.0 when vortices appear around the interface and shed into the flow. The interface rolls up as a result of the difference in the tangential velocity across the interface. Here, the roll ups take place on the sides and far from the middle of the domain. The central region is free from roll up as long as there is no jump in tangential velocity in these areas. The jump in tangential velocity across the interface with be discussed in the subsequent sections. 

         It should be pointed out that since the initial perturbation is symmetric, the flow is symmetric with respect to a vertical axis passing through half of the domain. However, these vortices develop on very small scales and little uneven existence will affect the vortex development. As the numerical method implies, these is always an uneven positioning of the front with respect to the stationary grid where the governing equations are solved. In other words, the moving grid or the front is not exactly positioned evenly relative to the Eulerian grid. Thus, the generation and development of the vortices do not occur evenly with respect to the axis of symmetry. This is visible in Fig~\ref{Fig:The flow structure}, and Fig.~\ref{Fig:9} at a later time. At time 1.98 the vortices appear and shed at uneven locations with respect to the axis of symmetry

     \begin{figure}[H]
	     	\centering
	     	\subfloat{\includegraphics[width=6cm]{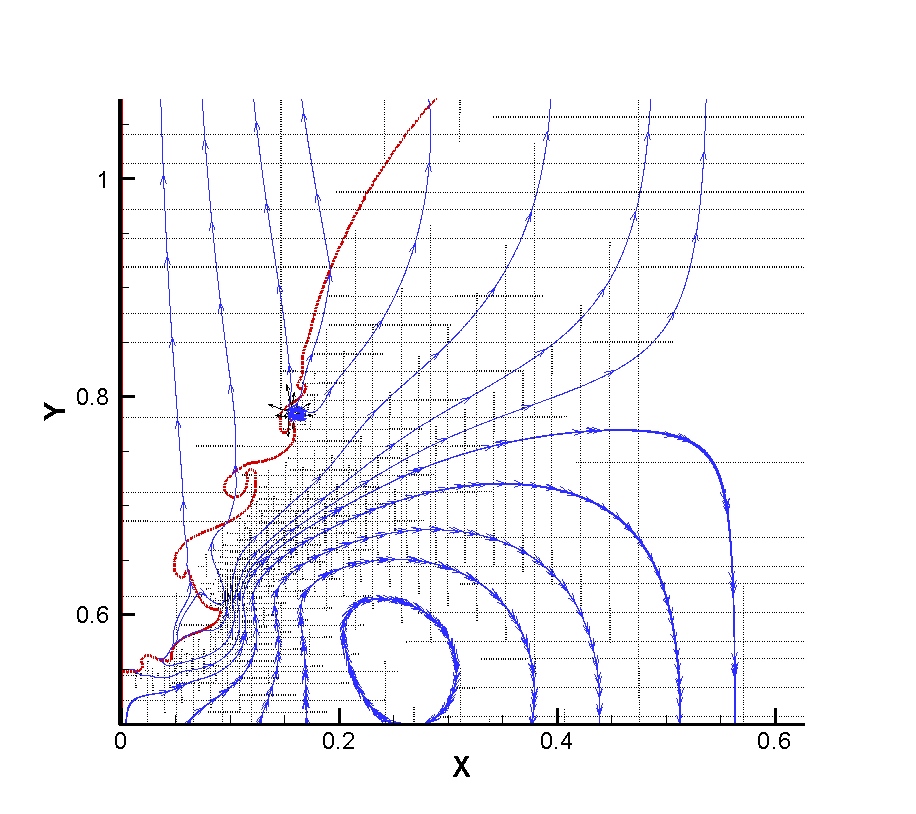}}
	     	\qquad
	     	\subfloat{\includegraphics[width=6cm]{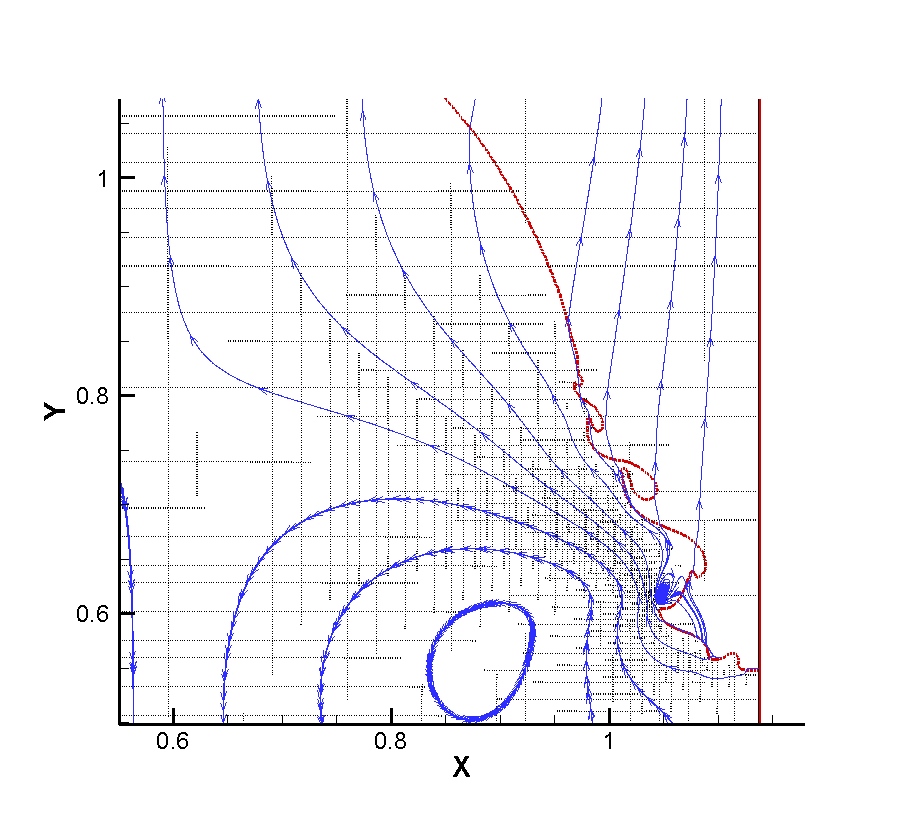}}
                 \qquad
                 \subfloat{\includegraphics[width=6cm]{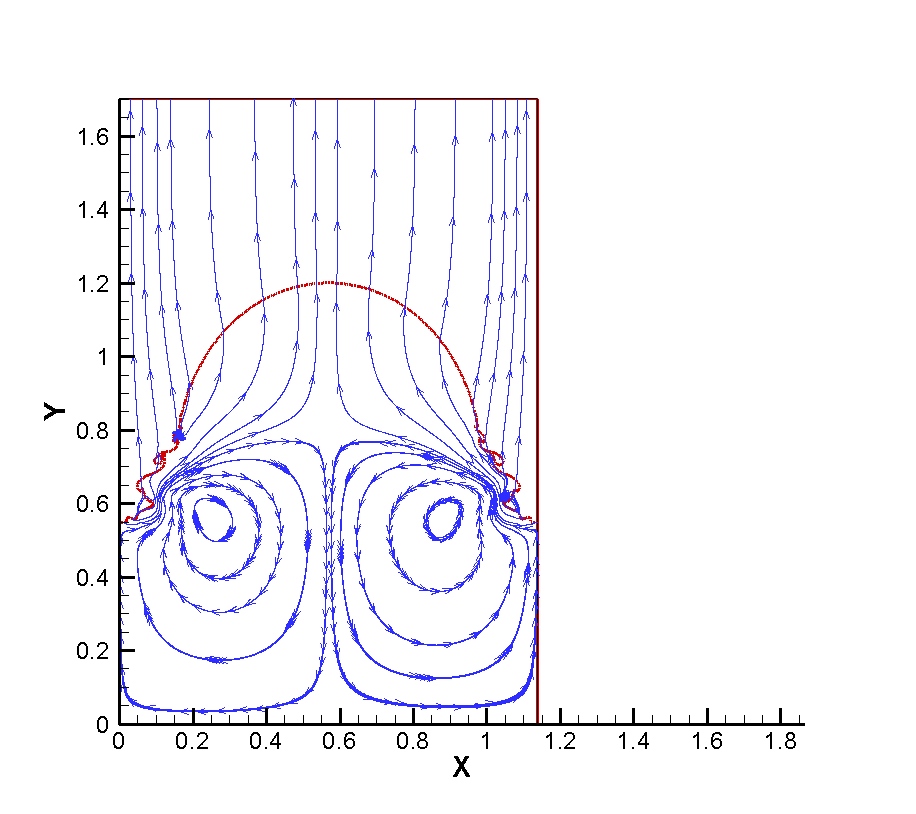}}
                 \qquad
                 \subfloat{\includegraphics[width=6cm]{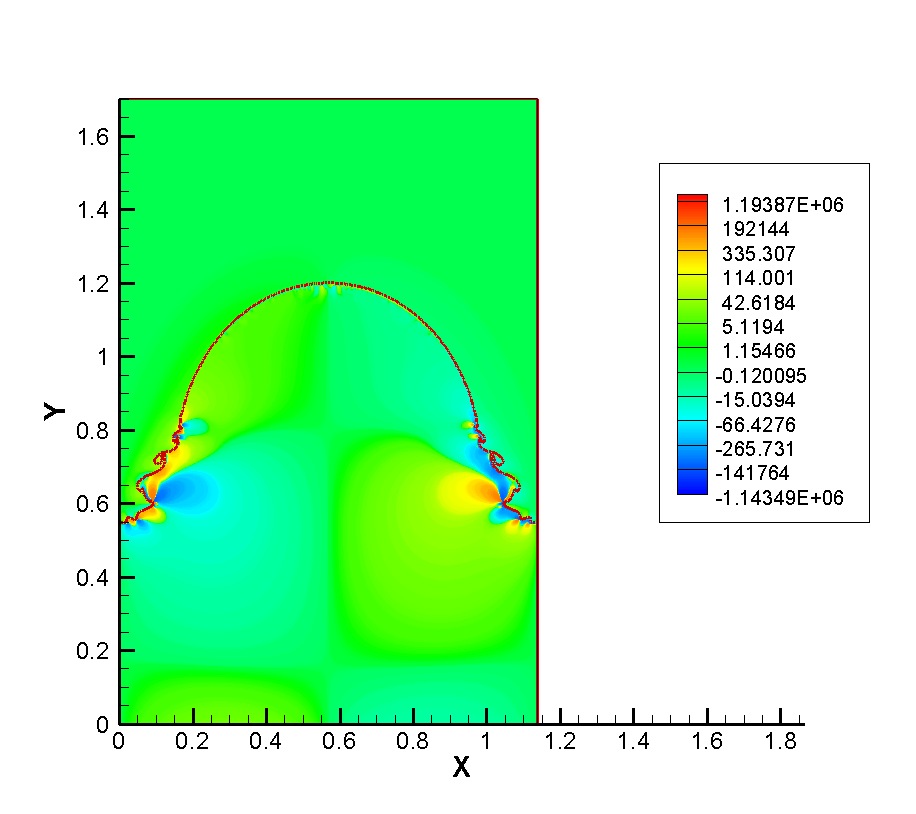}}
	     	\caption{The flow structure at time 1.98 for film boiling at a density ratio 1000. Typical vortices appear in the roll up regions in the left and right side of the front where the interface rolls up in these areas. The bottom right frame is a contour plot of vorticity. The grid resolution is $512\times768$.}
	     	\label{film boiling}
	     \end{figure}

\begin{figure}[H]
	     	\centering
	     	\subfloat{\includegraphics[width=6cm]{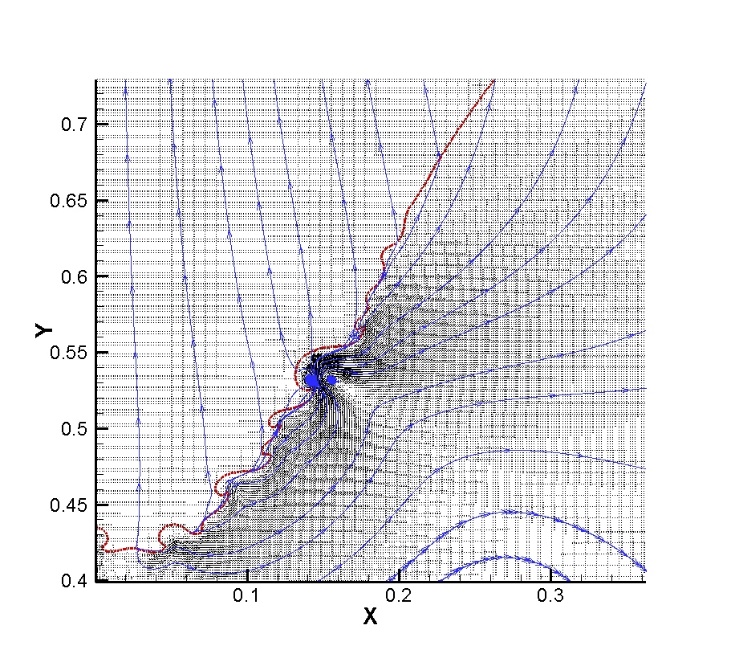}}
	     	\qquad
	     	\subfloat{\includegraphics[width=6cm]{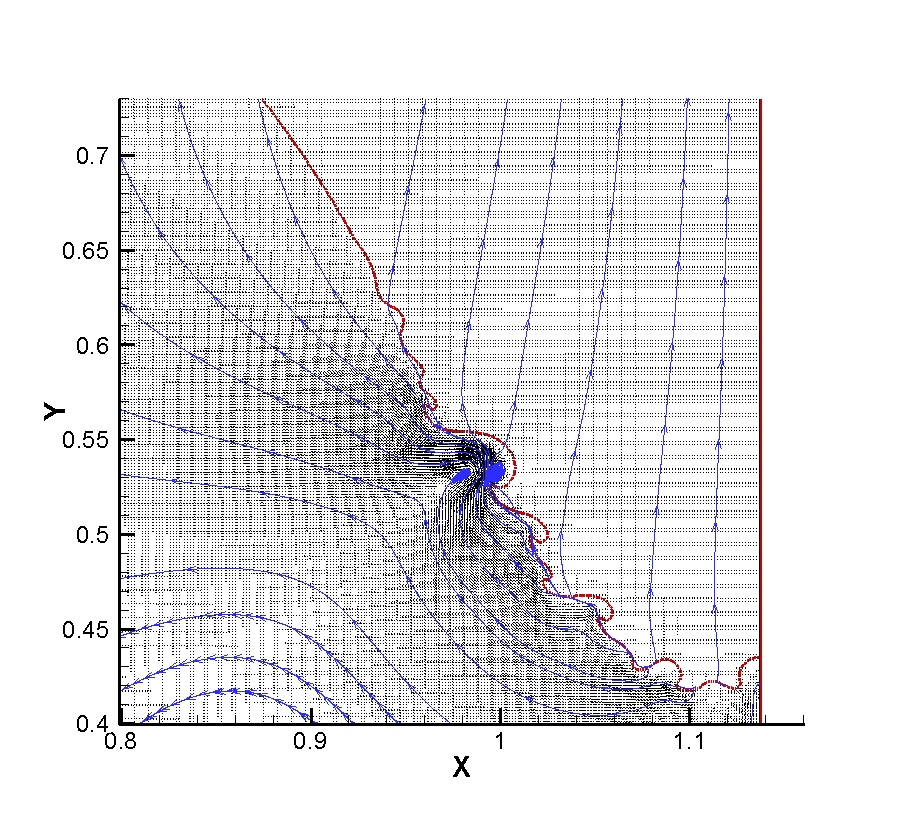}}
                 \qquad
                 \subfloat{\includegraphics[width=6cm]{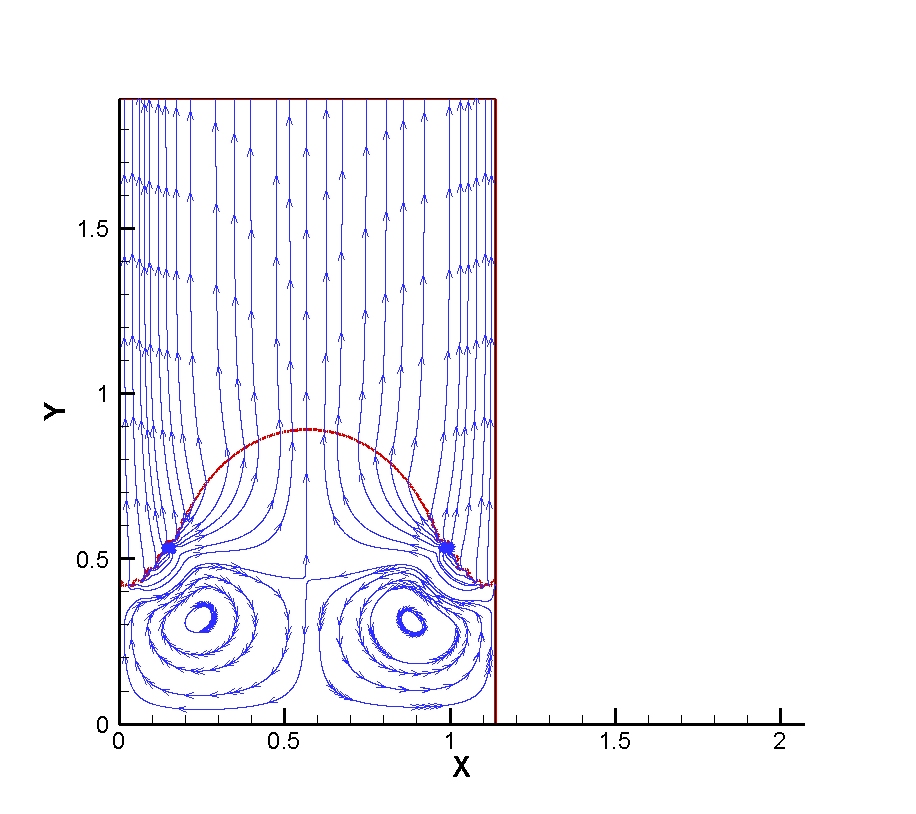}}
                 \qquad
                 \subfloat{\includegraphics[width=6cm]{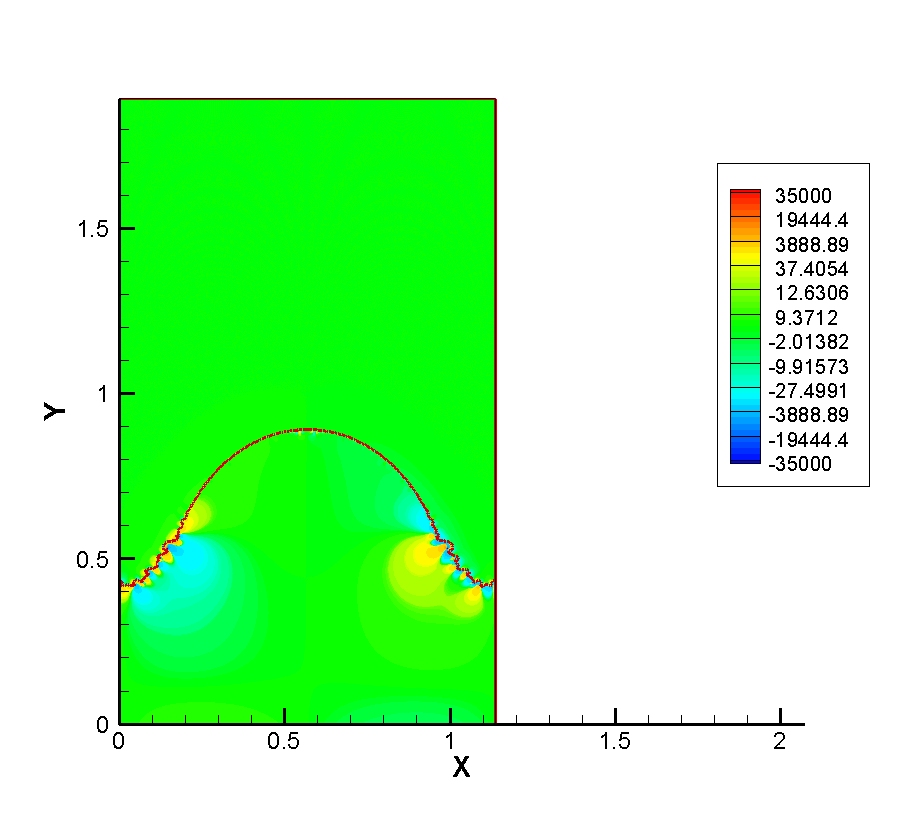}}
	     	\caption{The flow structure at time 1.27 for film boiling at a density ratio 1000. Typical vortices appear in the roll up regions in the left and right side of the front where the interface rolls up in these areas. The bottom right frame is a contour plot of vorticity. The grid resolution is $768\times1280$.}
           \label{Fig:The flow structure at time 1.27}
          \end{figure}

\begin{figure}[H]
	     	\centering
	     	\subfloat{\includegraphics[width=6cm]{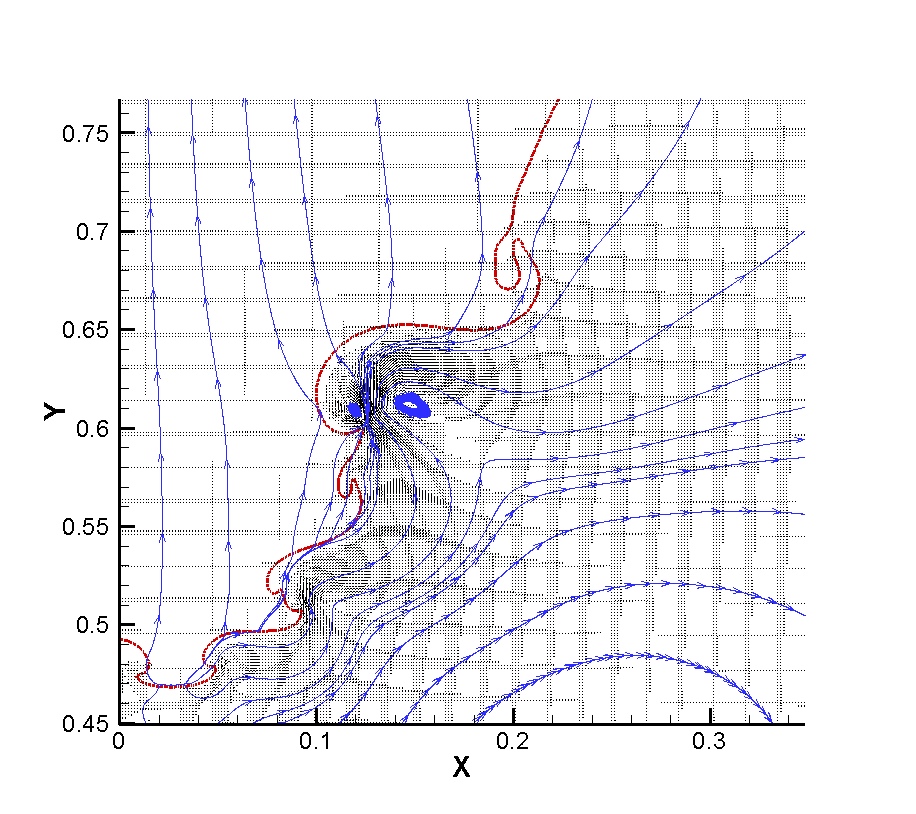}}
	     	\qquad
	     	\subfloat{\includegraphics[width=6cm]{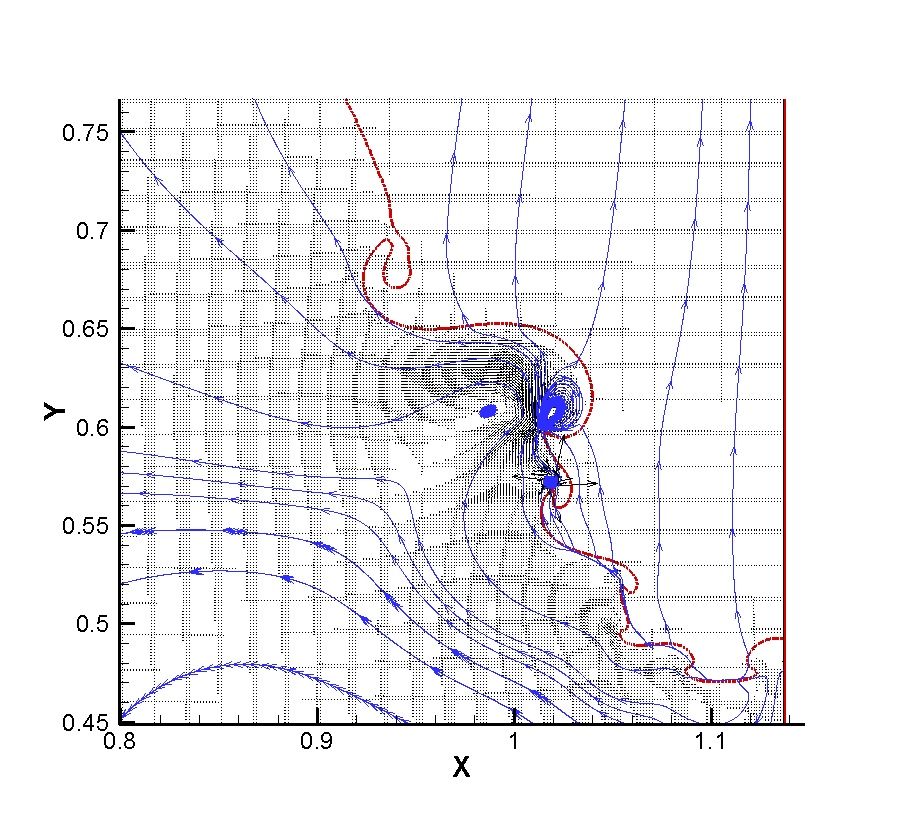}}
                 \qquad
                 \subfloat{\includegraphics[width=6cm]{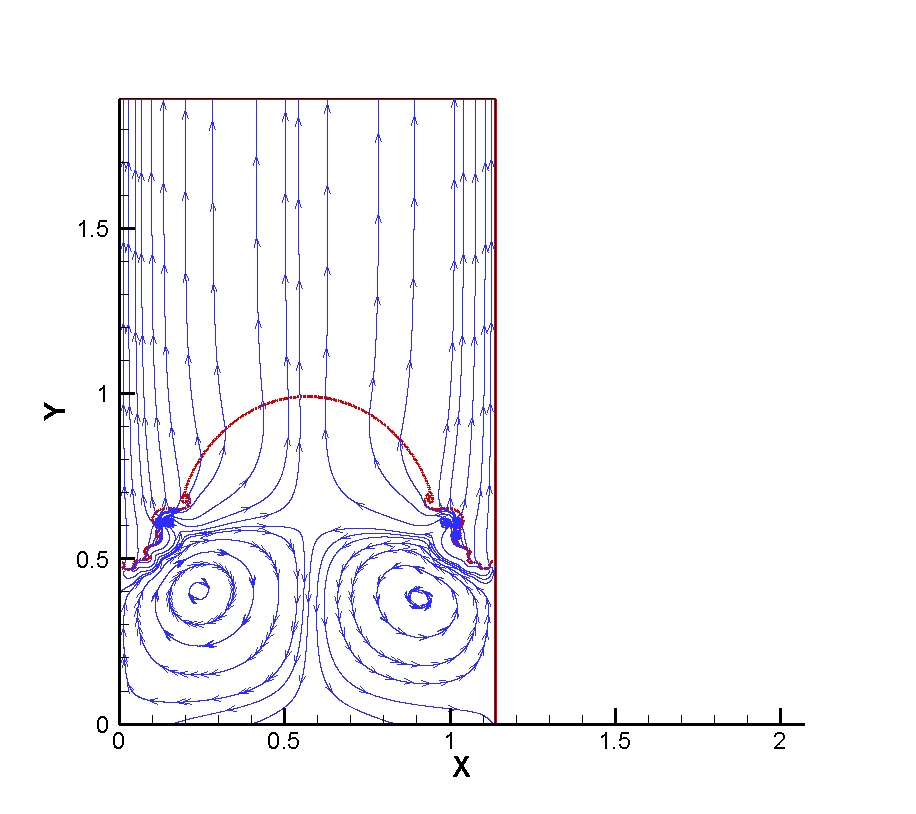}}
                 \qquad
                 \subfloat{\includegraphics[width=6cm]{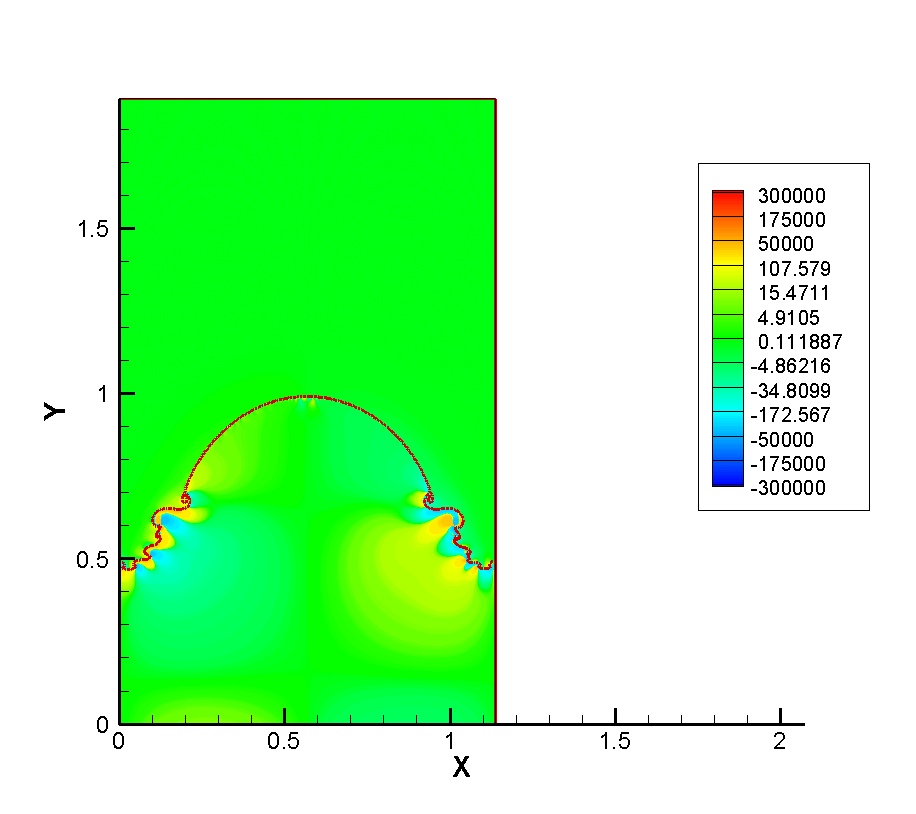}}
                 
	     	\caption{The flow structure at time 1.5 for film boiling at a density ratio 1000. Typical vortices appear in the roll up regions in the left and right side of the front where the interface rolls up in these areas. The bottom right frame is a contour plot of vorticity.The grid resolution is $768\times1280$.}
       \label{Fig:5}
             \end{figure}

           Since the vortices appear and shed on a small scale, they are influenced by little uneven positioning of the front relative to the stationary grid. The plot of vorticity fields at times 1.71 and 1.98 show an almost even pattern relative to the axis of symmetry except in the regions of roll up and vortex shedding. The streamlines also show little asymmetry which is due to an uneven vortex shedding in the roll up regions. These two figures show frequent development and shedding of vortices on the sides of the front where the roll up takes place. 
          Figure~\ref{Fig:The flow structure at time 1.27} and Figure~\ref{Fig:5} present the flow structure at times 1.27 and 1.5 at a higher grid resolution ($758\times1280$) respectively. The roll up of the front on the sides and vortex developments are visible in these areas similar to the low grid resolution. Nevertheless, the roll up is amplified here due to capturing finer structures at this grid resolution. 
          
       The circulation zones at the base of the flow are a little suppressed by the shedding of vortices on the sides of the front. This is visible in the streamlines plotted on the whole domain. The roll up is greatly amplified at a later time (1.5). The vorticity field shows that vorticity is concentrated in the roll up regions where vortex shedding in happening. Apart from vortex shedding, vorticity shows an almost symmetric pattern. Here, the fine structure shedding is captured better compared to the lower grid resolution. However, both grid resolutions are fine enough to capture vortices at small scales that shed into the flow (Figures~\ref{Fig:The flow structure} and~\ref{Fig:The flow structure at time 1.27}).
       
\begin{figure}[H]
	     	\centering
	     	\subfloat{\includegraphics[width=6cm]{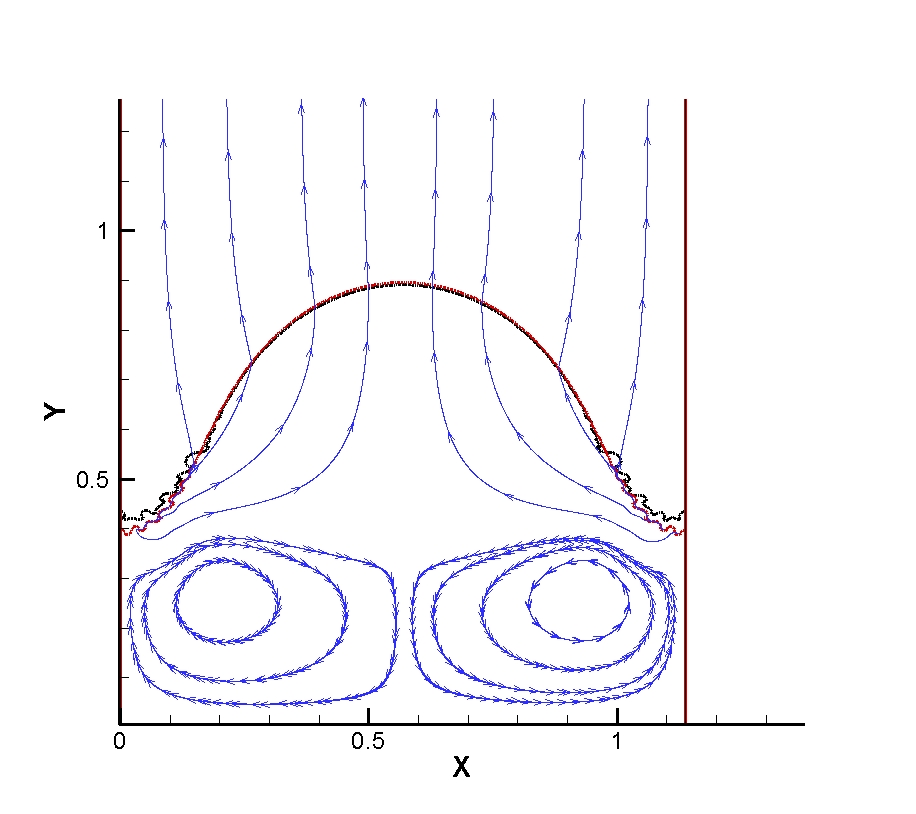}}
	     	\qquad
	     	\subfloat{\includegraphics[width=6cm]{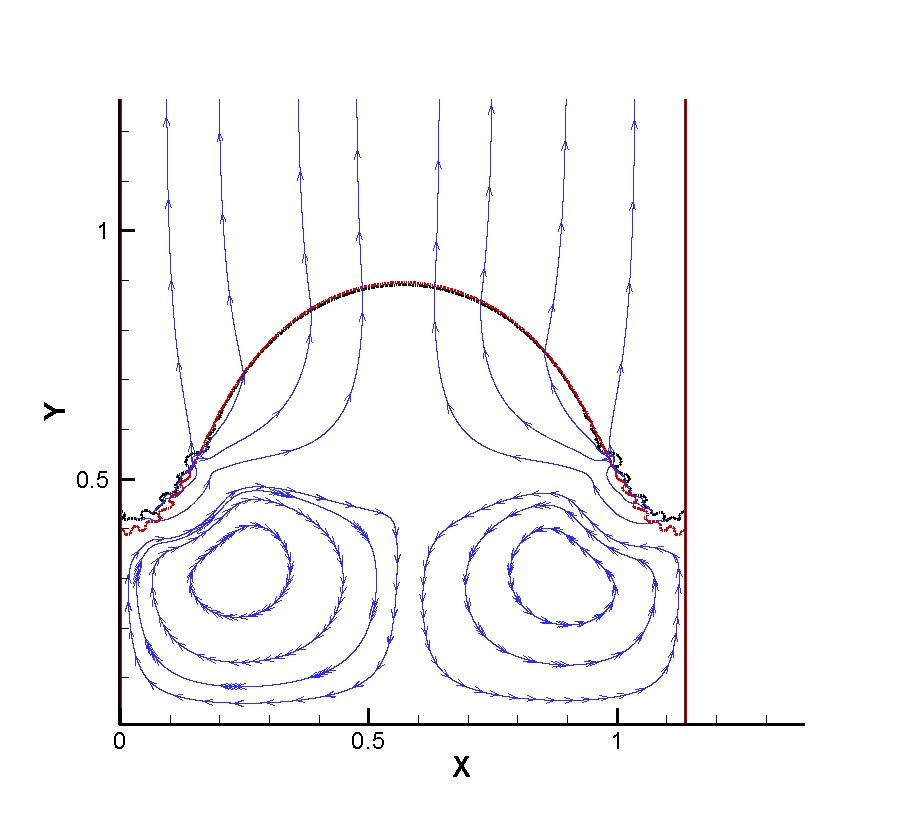}}
                 \qquad
                 
	     	\caption{Streamlines at two grid resolutions at time 1.27, the left frame is at $512\times768$ grids, while the right frame is for $768\times1280$ grid resolution. The red line and black line show the front position at low and high grid resolutions respectively.}
        \label{Fig:6}
             \end{figure}

\begin{figure}[H]
	     	\centering
	     	\subfloat{\includegraphics[width=7cm]{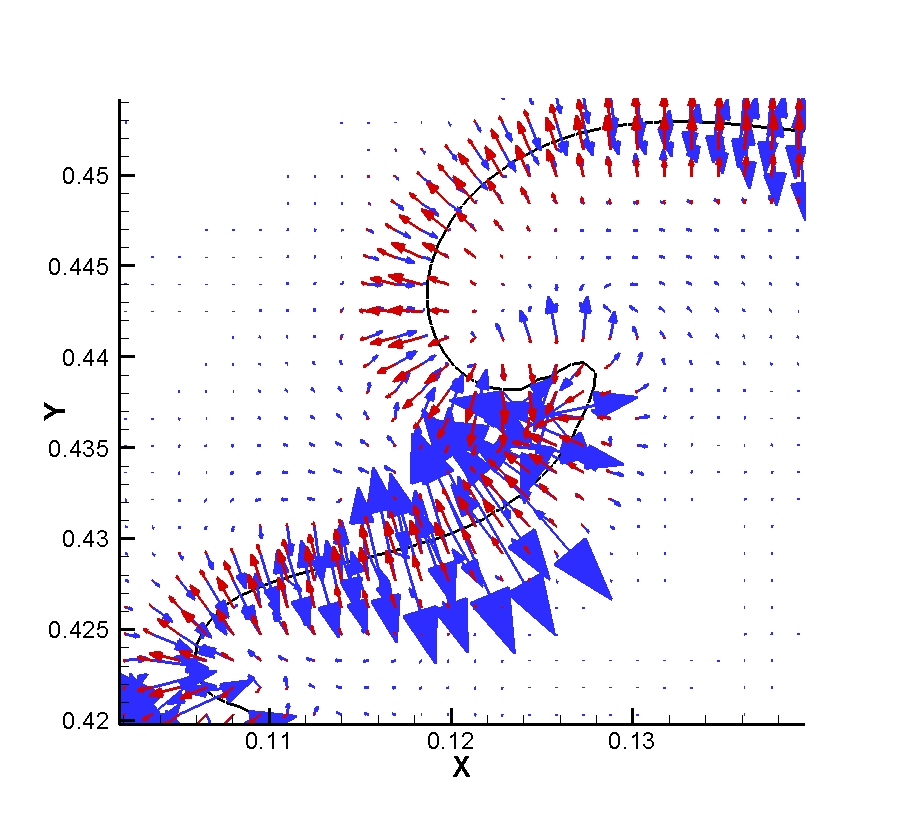}}
	     	\qquad
                 \label{a}
	     	\caption{Pressure gradient vector (blue arrows), and the density gradient vector (red arrows) in the roll up region along with the front. The density gradient vector is normal to the interface along the front. The grid resolution is $768\times1280$ and time is equal to 1.07.}
        \label{Fig:7}
             \end{figure}

\begin{figure}[H]
	     	\centering
	     	\subfloat{\includegraphics[width=6cm]{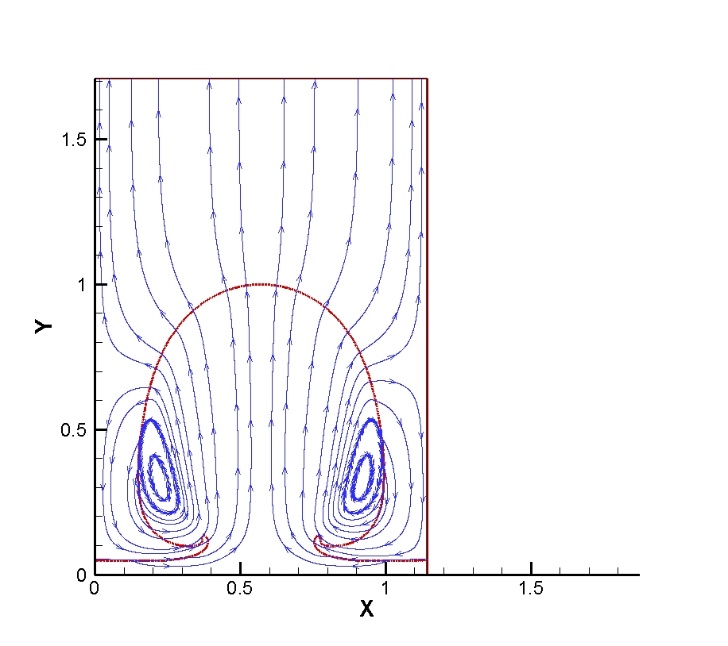}}
	     	\qquad
	     	\subfloat{\includegraphics[width=6cm]{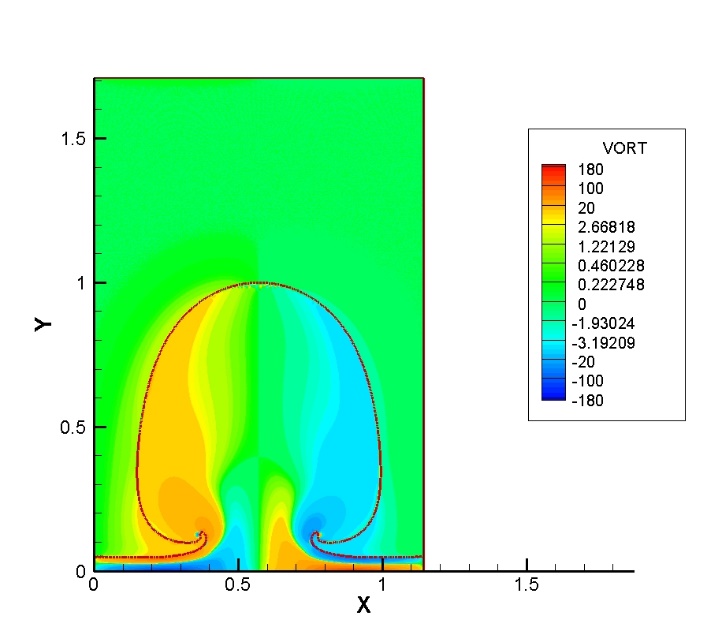}}
                 \qquad
                 \label{a}
	     	\caption{The flow structure at time 8.64 for film boiling at a density ratio100. The interface is quite smooth at this density ratio. The right frame is a contour plot of vorticity.The grid resolution is $512\times768$.}
        \label{Fig:8}
             \end{figure}

             Simulations at these two grid resolutions are compared in Fig.~\ref{Fig:6}. Here the front position and streamlines are compared at time 1.27. The left frame shows streamlines at $512\times768$ grid resolution, and the right frame shows streamlines at higher grid resolution ($768\times1280$). The suppression of the streamlines at low grid resolution is due to a lower front position on the sides (red line) compared to the higher grid resolution (black line). The front positions are almost the same in other regions. Also, the interface roll up is amplified at the higher grid resolution. Streamlines show similar patterns away from the circulation zones.
                      The vortex development and shedding is elaborated more by addressing the baroclinic term in Fig.~\ref{Fig:7}. The pressure gradient vector (blue arrow) and the density gradient vector (red arrow) are plotted in the roll up region including the interface shape. The grid resolution is $768\times1280$ and time is equal to $1.07$. Here, the density gradient vector is naturally normal to the interface which separates two density fields. Thus, the baroclinic term is nonzero whenever the pressure gradient vector is not normal to the interface. The pressure gradient on the interface is due to gravity, interfacial tension and the liquid evaporation which develops a nonzero divergence at the interface. It is visible that the pressure gradient vector is inclined with respect to the interface in the roll up region where a narrow neck is formed. It is postulated that this term is responsible for the vortex development and shedding in small scales on the interface. This term is large at high density ratios (1000) where the simulations are performed. The behavior is not seen at low density ratio (100) (Fig.~\ref{Fig:8}). 
                      
\begin{figure}[H]
	     	\centering
	     	\subfloat{\includegraphics[width=8cm]{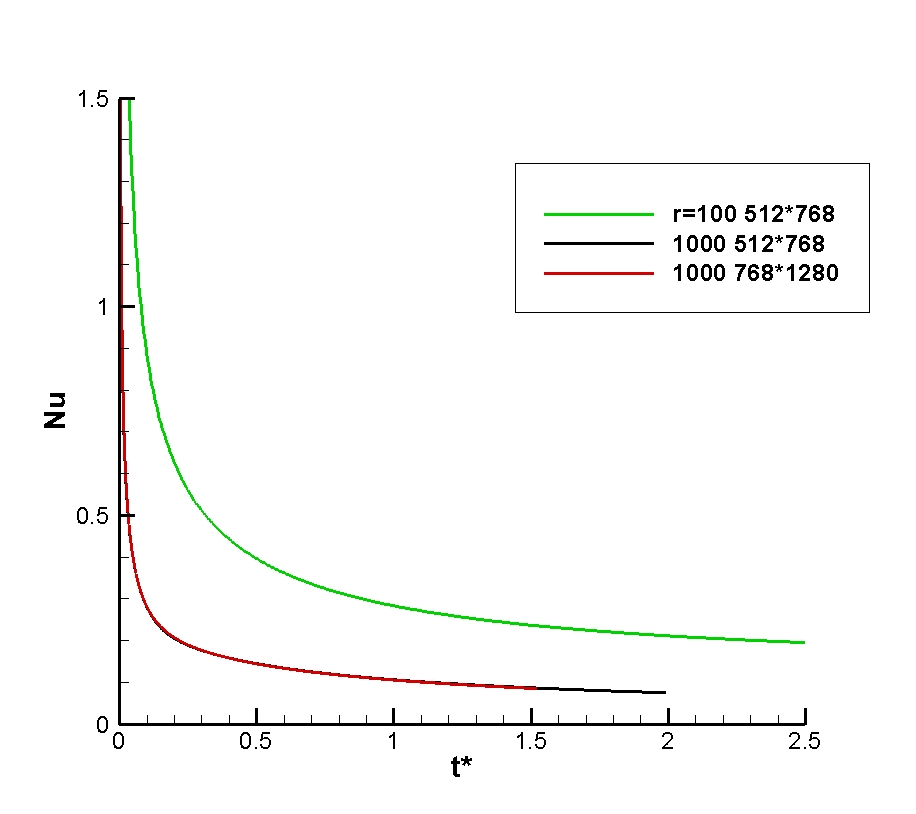}}
	     	\qquad
                 
	     	\caption{The special averaged Nusselt number versus time for two density ratios (100 and 1000). The Nusselt number is a little lower than experimental correlations.}
        \label{Fig:10}
             \end{figure}

                    The flow structure is plotted for density ratio 100 at time 8.64 in Fig.~\ref{Fig:8} for comparison. The streamlines show circulation zones on the sides of the interface. The neck is relatively narrower compared to the higher density ratio (1000). The interface is fairly smooth with a small neck at the base. Streamlines are quite symmetric in this case. The vorticity field also shows quite smooth variations with a symmetric pattern. Here the enstrophy is relatively small and smooth compared to the high density ratio (Fig.~\ref{Fig:9}). There is no vortex shedding in this case. 
             Enstrophy shows a better understanding of the flow development as time proceeds (Fig.~\ref{Fig:9}). Here the ensttrophy is made dimensionless using the characteristic time associated with the problem. The enstrophy shows the same trend for two grid resolutions up to time 1.0. At this stage the high grid resolution simulation starts departing from the low grid resolution. Small scale structures are captured better at high grid resolution. As a result, the vortex development occurs at a faster rate at higher grid resolution. The shedding of vortices is amplified and develops to a higher extend for this case. The picks that appear in the plot are due to vortex shedding in the flow. It should be emphasized that here the enstrophies are almost the same up to time 1.0 which can be justified as a grid convergence for the problem. The departure from each other is due to capturing the fine structures more accurately at higher grid resolution. A detailed grid convergence study was given in Fig.~\ref{Fig:6}. 
             A case with a lower density ratio (100) is also shown for comparison. Enstrophy is much smaller for this case compared to the high density ratio. It is an order of magnitude smaller for this case. There is no peak and vortex shedding in this case. The flow is smooth without any interface roll up as is visible in the flow structure for this case (Fig.~\ref{Fig:8}).
             The spatial average Nusselt numbers are plotted versus time in Fig.~\ref{Fig:10}. The Nusselt number is relatively higher at lower density ratio (100). The Nusselt numbers are almost the same at two grid resolutions for high density ratio (1000). Table \ref{tab2} compares the average Nusselt numbers with experimental correlations. The average Nusselt numbers predicted here are a little lower than experimental correlations. These are nearly the same as those obtained by Mortazavi \cite{mortazavi2022toward} using numerical simulations at high density ratios. It should be pointed out that here simulations have not been performed for long time to capture the release of periodic bubbles, and the intention is not to do such simulations. Thus, there may be some approximation in obtaining the average Nusselt number from the current simulations (see Mortazavi \cite{mortazavi2022toward}). 

             \begin{table}[h]
            \caption{Comparing the Nusselt number from present study with experimental correlations}\label{tab2}%
             \begin{tabular}{@{}llll@{}}
             \toprule
               & $r$ = 100  & $r$ = 1000 \\
         \midrule
           present study     & 0.2   & 0.08    \\
        Hamil and Baumeister    & 0.52   & 0.29    \\
        Berenson    & 0.46   & 0.26    \\
        Chang    & 0.37   & 0.17    \\
        
        Klimenko and Shelepen    & 0.3   & 0.14   \\
         \botrule
       \end{tabular}
       
       \end{table}

        It should be pointed out that here the Nusselt number depends on the dimensionless heat flux at the wall as is predicted by Mortazavi [6]. Higher Nusselt number are obtained when the wall heat flux is reduced. Experimental correlations are not sensitive to the wall heat flux. Agreement gets better at lower wall heat fluxes as predicted by Mortazavi \cite{mortazavi2022toward}. 
             A stability analysis is performed on the flow based on the linear stability theory. Here the effect of phase change is ignored. There is no analytic solution when phase change occurs. However, the effect of phase change can be addressed by looking at the characteristic time associated with it and comparing with the time scale associated with Kelvin-Helmholtz instability. A rough estimate for these two time scales was performed in the simulations and showed that the phase change characteristic time is relatively smaller than the characteristic time related to Kelvin-Helmholtz instability. The time scale related to phase change is calculated from 
             $t_{\mathsf{phase-change}}=\sfrac{\lambda_{d_{2}}}{(\sfrac{\dot{m}}{\rho_{\ell}}})$
             where $\dot{m}$  is the evaporation rate. The time scale associated with the Kelvin-Helmholtz instability is obtained from 
             $t_{k-h}=\sfrac{\lambda_{d_{2}}}{(U_{\ell}-\ U_g\ })$
             where $U_{\ell}$ and $U_g$ are the liquid and gas velocities across the interface. Numerical values are obtained as 0.526 and 1.653 respectively in dimensionless form. As a result, ignoring the effect of phase change is a relatively reasonable approximation. The stability analysis includes the term due to the jump in the tangential velocity (Kelvin-Helmholtz instability), the term due to the jump in the density across the interface (Rayleigh-Taylor instability), and also the term due to capillary instability. The complex growth rate can be derived from linear theory which is as follows:
             
              \begin{align}
 s = -i k \frac{\rho_1 U_1 + \rho_2 U_2}{\rho_1 + \rho_2} \pm \sqrt{\frac{k^2 \rho_1 \rho_2 (U_1 - U_2)^2}{(\rho_1 + \rho_2)^2} - \frac{\gamma k^3}{\rho_1 + \rho_2} - \frac{k g (\rho_1 - \rho_2)}{\rho_1 + \rho_2}}
             \end{align}
               Where the subscript 1 and 2 refer to gas and liquid respectively. k is the wave number $(\frac{2\pi}{\lambda})$ and $\gamma$ is the interfacial tension. The imaginary part of this expression is the phase velocity and the real part is the growth rate:
               
               \begin{align}
               \sigma= \sqrt{\frac{k^2\rho_{1}\rho_{2}{( U_1-U_2)}^{2}}{{(\rho_1+ \rho_2)}^2}- \frac{\gamma k^3}{\rho_1+\rho_2}-\frac{k g(\rho_1-\rho_2)}{\rho_1+ \rho_2}}
               \end{align}
               
               The flow is unstable to small perturbations if the growth rate is real. The growth rate in dimensionless form is: 
               \begin{align}
               \sigma^{\star}=\frac{\sigma}{\ kU_{c}}=\frac{1}{U_{c}} \sqrt{\frac{\rho_{1}\rho_{2}{( U_1-U_2)}^{2}}{{(\rho_1+ \rho_2)}^{2}}-\frac{\gamma k}{\rho_{1}+\rho_{2}}- \frac{g (\rho_1-\rho_2)}{k(\rho_1+\rho_2)}}
               \end{align}
               Where $U_c$ is a characteristic velocity defined by $U_c=\left(\lambda_{d_{2}}g\right)^{\sfrac{1}{2}}$.

\begin{figure}[H]
	     	\centering
	     	\subfloat{\includegraphics[width=6cm]{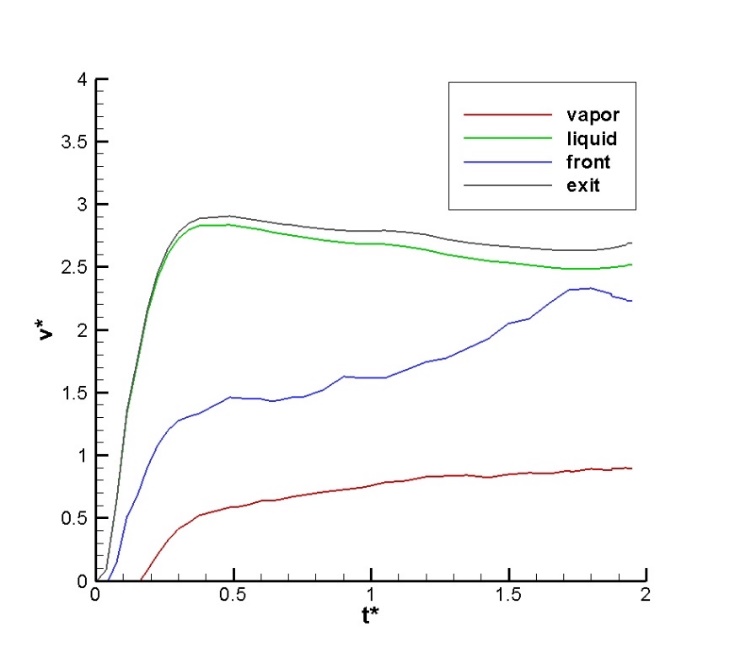}}
	     	\qquad
	     	\subfloat{\includegraphics[width=6cm]{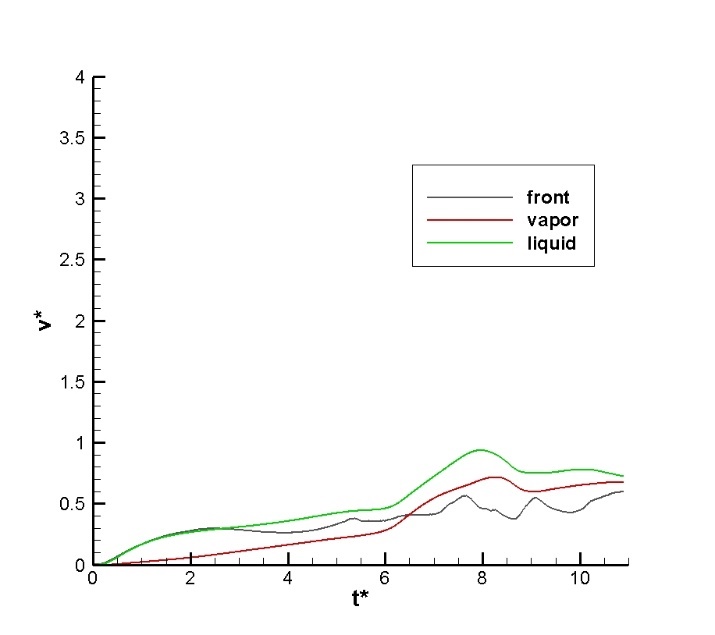}}
                 \qquad
                 
	     	\caption{Average vapor, liquid, interface and exit velocity at the upper boundary versus time for film boiling at a high density ratio (1000) (right frame), same velocities at density ratio 100 (left frame).}
        \label{Fig:11}
             \end{figure}

        \noindent In order to make an estimate of the growth rate in the current problem, the average gas velocity and liquid velocity are calculated and plotted with respect to time in Fig.~\ref{Fig:11} for two density ratios (100,1000). The average front velocity and the velocity at the exit (upper) boundary are also plotted for comparison. The liquid velocity is obtained by multiplying the velocity by the indicator function at every grid point, and averaging it over the grid points that locate in the liquid phase. Here, the v-component of the velocity is chosen for averaging, as the average of the u-component is zero due to symmetry with respect to the central axis. The indicator function is one in the liquid phase and zero in the gas phase. The gas velocity is also calculated in a similar way by multiplying the velocity by one minus the indicator function. The average liquid velocity is nearly the same as the exit velocity as it should be. The velocity is normalized by the characteristic velocity: $U_c=\left(\lambda_{d_{2}} g\right)^{\sfrac{1}{2}}$.
              
             The front velocity lies between the gas velocity and the liquid velocity for high density ratio. The front velocity does not have a smooth variation due to an irregular shape of the interface. The velocities are much smaller for the lower density ratio. Initially the front velocity is between the gas velocity and the liquid velocity at low density ratio since the interface is growing. However, the base of the front gets stable close to the heated wall as the bubble grows (Fig.~\ref{Fig:8}). This is in contrast to the high density ratio (1000) where the whole bubble moves upward, and does not develop a stable base. As a result, the front velocity stays below the gas velocity at later stages. In other words, since the base of the bubble gets stable close to the heated wall, the interface velocity does not grow substantially compared to the gas velocity (Fig.~\ref{Fig:8}). This is not the case for the high density ratio where the bubble does not have a stable base. The front velocity also does not have a smooth variation since the front includes different regions of the flow where the velocity may increase or decrease locally according to the flow configuration. The difference between the gas velocity and the liquid velocity is roughly an order of magnitude smaller for the low density ratio (100), compared to the high density ratio (1000). This in fact affects the Kelvin-Helmholtz instability associated with it for two cases. To be more precise, the roll up which takes place as a result of the jump in tangential velocity across the interface does not occur at low density ratio (100). Thus, it is natural to observe a smooth interface without any roll up at low density ratio (100). The bubble has almost reached the upper boundary close to the end of simulation for low density ratio (100) (not plotted). Kelvin-Helmholtz instability occurs at high density ratio (1000) due to a large difference between that liquid and the gas velocity across the interface (the jump in the tangential velocity across the interface). Hence, Kelvin-Helmholtz is very susceptible at high density ratios as has been captured in the current simulations at a high density ratio. 
             Table \ref{tab3} presents the dimensionless growth rate for two density ratios based on the linear theory. The gas density is raised by a factor of ten for the lower density ratio. Other parameters are the same. The difference between the liquid velocity and the gas velocity across the interface is extracted from Fig.~\ref{Fig:11} for two density ratios. A rough averaging over time is used to obtain the velocity differences. The wave number was obtained based on the domain length which is two times the most unstable wavelength of Rayleigh-Taylor instability. It can be seen that the growth rate is positive for both cases, so as expected the interface grows for small perturbations. The growth rates are almost the same for two density ratios examined, but a little higher at larger density ratio. 
         \begin{table}[h]
          \caption{Dimensionless growth rate for two density ratios }\label{tab3}
       \begin{tabular*}{\textwidth}{@{\extracolsep\fill}lcccccc}
       \toprule%
         $r$ & $\lambda_{d_{2}}$ & $k=\frac{2\pi}{2\lambda_{d_{2}}}$ & $U_c$ & ${\Delta U}^\star= U_{\ell}^\star-\ U_g^\star$ & $t_c$ & $\sigma^\star$ \\
         \midrule
        100  & 0.57127 & 5.496 & 0.5344 & 2.0 & 0.2668 & 0.535\\
         1000   & 0.5684 & 5.52  & 0.533  & 0.2 & 0.2668 & 0.544\\
         \botrule
          \end{tabular*}
        
        \end{table}
        
       Both cases are subject to instability, i.e. the interface is unstable to small perturbations, however, the contribution from Kelvin-Helmholtz instability (roll up of the interface) is much smaller at low density ratio (100) compared to high density ratio (1000). Evidently, the interface roll up is not seen at low density ratio (the jump in tangential velocity is much smaller). On the other hand, the interface grows, and the vapor bubble develops and moves upward in both cases due to the Rayleigh-Taylor instability.
       \section{Conclusions }\label{sec6}
        Film boiling at high density ratio (1000) was studied by numerical simulations with high grid resolution. The interface rolls up in small scales at high density ratio due to large difference between the gas and liquid. It does not occur at a relatively low density ratio (100) where the jump in the velocity across the interface is an order of magnitude smaller. Also, the roll up takes place at relatively high grid resolution (512 grids per width of the domain). The roll up in small scales amplifies as the grid is refined. The enstrophy inside the flow shows that it is relatively smooth at initial stages, and is almost equal at two high grid resolutions. The enstrophy eventually becomes spiky due to vortex shedding in the gas phase. The enstrophies depart from each other for two grid resolutions at certain time ($t^{\star}=1.0$). This is mainly due to capturing very small scales at the higher grid resolution. It turns out that vertex shedding happens in small scales that are resolved better at higher grid resolution. The author postulates that the development of vortices around the interface is due to the baroclinic term in the vorticity transport equation. This term gets large at high density ratios (1000) around the interface. The pressure gradient around the interface is mainly due to liquid evaporation and the interfacial tension. This term also gets large in the roll up region where the interface area and curvature increase. The increase in the area raises the evaporation rate locally that also amplifies the pressure gradient. A linear stability analysis shows that the growth rate is positive for two density ratios examined (100,1000), and is a little higher at the larger density ratio. The Nusselt numbers obtained are a little lower than experimental correlations consistent with previous work by Mortazavi \cite{mortazavi2022toward}. Future work includes simulations in three dimensions at high density ratio with the help of clusters of computers and parallel computing.
        \section*{Declaration of interests}
        
       The authors declare that they have no known competing financial interests or personal relationships that could have appeared to influence the work reported in this paper.
       
       \bmhead{Acknowledgement}
       
       The authors gratefully acknowledge the Sheikh Bahaei National High Performance Computing Center (SBNHPCC) for providing computing facilities and time. SBNHPCC is supported by scientific and technological department of presidential office and Isfahan University of Technology (IUT). 


        \bibliography{sn-bibliography}

      \end{document}